\documentclass[twocolumn,showpacs,prl]{revtex4}

\usepackage{amssymb,amsmath}
\usepackage{graphicx}
\usepackage{xcolor}

\newcommand{\ket}[1]{|#1\rangle}

\newcommand{\eq}{\begin{equation}}
\newcommand{\fine}{\end{equation}}
\newcommand{\el}{\ell}
\newcommand{\kk}{{\bf k}}
\newcommand{\cluster}{\ket{\Phi^{\text{lin}}_4}}

\begin{document}

\title{Active one-way quantum computation with 2-photon 4-qubit cluster states }
\author{Giuseppe Vallone}
\homepage{http://quantumoptics.phys.uniroma1.it/}
\author{Enrico Pomarico}
\altaffiliation[Present address: ]{Univ. de Gen\`eve
GAP-Optique,
Rue de l'\'Ecole-de-M�decine 20,
CH-1211 Gen\`eve 4,
Suisse}
\author{Francesco De Martini}
\homepage{http://quantumoptics.phys.uniroma1.it/}
\author{Paolo Mataloni}
\homepage{http://quantumoptics.phys.uniroma1.it/}
\affiliation{
Dipartimento di Fisica dell'Universit\'{a} ``La Sapienza'' and
Consorzio Nazionale Interuniversitario per le Scienze Fisiche della Materia,
Roma, 00185 Italy}

\begin{abstract}
{By using 2-photon 4-qubit cluster states we demonstrate deterministic one-way quantum computation in single 
qubit rotation algorithm. In this operation
feed-forward measurements are automatically implemented by properly choosing the measurement basis of the qubits, while Pauli 
error corrections are realized by using two fast driven Pockels cells. 
We realized also a C-NOT gate for equatorial 
qubits and a C-Phase gate for a generic target qubit. Our results 
demonstrate that 2-photon cluster states can be used
for rapid and efficient deterministic one-way quantum computing.}
\end{abstract}


\maketitle




Cluster states are the basic resource for one-way quantum computation (QC) \cite{01-bri-aon}. 
In the standard QC approach any quantum algorithm can be
realized by a sequence of single qubit rotations and two qubit gates, such
as C-NOT and C-Phase \cite{01-kni-asc}. Deterministic one-way QC is based on
the initial preparation of entangled qubits in a cluster state, followed by
a temporally ordered pattern of single qubit measurements and feed-forward (FF)
operations depending on the outcome of the already
measured qubits \cite{01-bri-aon}. We can consider two different types of FF operations: 
\textit{i)} the intermediate feed-forward measurements, i.e. the choice of the measurement basis depending
on the previous measurement outcomes and
\textit{ii)} the Pauli matrix feed-forward corrections on the
final output state. Two qubit gates can be realized by exploiting the
existing entanglement between qubits. In this way the difficulties of
standard QC, related to the implementation of two qubit gates, are
transferred to the preparation of the state.

One-way QC was experimentally realized by using 4-photon cluster states
\cite{05-wal-exp} and, later, by implementing Pauli error corrections with 
active feed-forward \cite{07-pre-hig}. Very recently, a proof of principle of one-way QC
(namely, the Grover's search algorithm and a particular C-Phase gate) 
was given by employing 2-photons entangled in a 4-qubit cluster state, without active
feed-forward \cite{07-che-exp}. 
The possibility of encoding more qubits on the same photon determines
important advantages. For instance the overall detection efficiency $\eta$, and then the
detection rate, is constant. In fact, in general it scales as $\eta ^{N}$, being $N$
the number of photons. Moreover 2-photon cluster states, realized by entangling the 
polarization ($\pi $) and linear momentum (${\mathbf{k}}$) degrees of freedom, 
outperform the earlier results obtained by 4-photon cluster
states since the state fidelities can be far larger and the count rate can
be higher by 3 - 4 orders of magnitude \cite{07-val-rea,07-che-exp}.  

One may ask how feed-foward operations can be implemented in the case of 2-photon multiqubit cluster states
and if the different degrees of freedom of the photons are computationally equivalent. 
In this Letter we demonstrate that a simple smart choice of the measurements basis
can be used in this case to deterministically perform the single qubit rotation algorithm, 
instead of adopting intermediate active feed-forward measurements (type \textit{i)}), 
which are necessary in the case of QC with 4-photon cluster states 
\cite{07-pre-hig}. Moreover feed-forward corrections (type \textit{ii)}) 
have been implemented in the same algorithm by using active fast driven 
electro-optics modulators. In this way, high speed deterministic one-way single qubit rotations 
are realized with 2-photon 4-qubit cluster states 
\footnote{Here ``deterministic'' refers to the success of the one-way algorithm and not to the overall process beacuse of 
the random generation of SPDC pairs.}.
In the same context, we have also verified the equivalence existing 
between the degrees of freedom of polarization and linear momentum 
by using either $\mathbf{k}$ or $\pi $ as QC output qubit. 

\begin{figure}[t]
\begin{center}
\includegraphics[width=8cm]{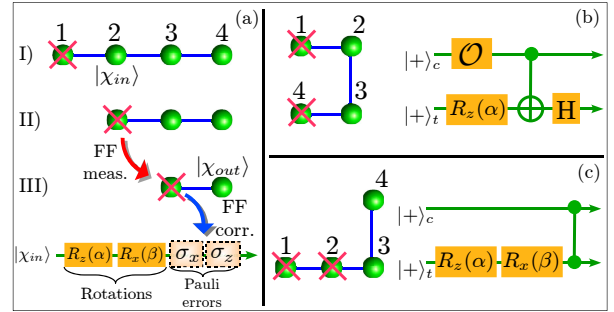}
\end{center}
\caption{(a) Top: measurement pattern for arbitrary single qubit rotations on a 4-qubit
linear cluster state, carried out in three steps (I, II, III). In each
measurement, indicated by a red cross, the information travels from left to
right. Feed-forward measurements and corrections are respectively
indicated by the red and blue arrows. Bottom: equivalent logical circuit. 
(b) C-NOT realization via measurement of qubits 
1, 4 on the horseshoe cluster (left) and equivalent circuit (right). 
(c) Universal C-Phase gate realization via measurement of qubits 1 and 2 
(left) and equivalent circuit
(right).}
\label{fig:cluster}
\end{figure}

Besides single qubit rotations, two qubit gates, such as the C-NOT gate, 
are required for the realization of arbitrary quantum algorithms \cite{01-kni-asc}. 
The realization of a C-NOT gate for equatorial qubits and a universal C-Phase gate acting on arbitrary
target qubits is also presented in the present Letter. 

In our experiment 2-photon 4-qubit cluster states were generated by using the
methods described in \cite{07-val-rea,05-bar-pol,06-bar-enh,05-cin-all}, to which we refer for details. Two
photons belonging to the cluster state \footnote{%
The state \eqref{cluster} is equivalent to that generated in \cite
{07-val-rea} up to single qubit transformations.} 
\begin{equation}
\begin{aligned}\label{cluster} |C_{4}\rangle=&\frac{1}{2}(|H\ell \rangle
_{A}|Hr\rangle _{B}- |Hr\rangle _{A}|H\ell \rangle_{B}\\ &+|Vr\rangle
_{A}|V\ell \rangle _{B}+|V\ell \rangle _{A}|Vr\rangle_{B}) \end{aligned}
\end{equation}
are generated either with horizontal ($H$) or vertical ($V$) polarization,
in the left ($\ell $) or right ($r$) mode of the Alice ($A$) or Bob ($B$)
side (see Fig. 1 of \cite{07-val-rea}). 
Cluster states were detected with fidelity $F=0.880\pm 0.013$, 
as obtained from the measurement of the stabilizer operators of $|C_{4}\rangle $ \cite{05-kie-exp}.
\begin{figure}[t]
\centering
\includegraphics[width=8cm]{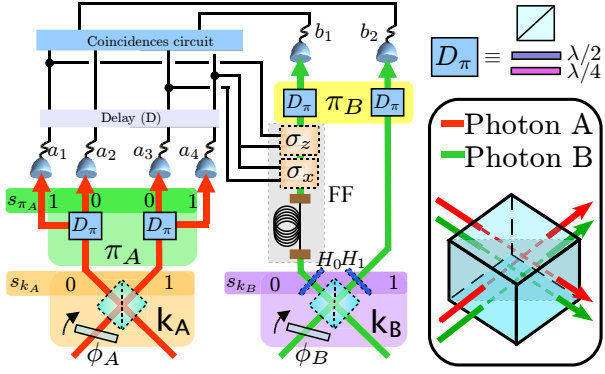}
\caption{Measurement setup for photons A and B. Momentum qubits $\mathbf{k}%
_{A}$ and $\mathbf{k}_{B}$ are measured by two thin glass plates ($\phi _{A}
$, $\phi _{B}$), acting as phase shifters and inserted before the BS. 
Polarization qubits $\pi _{A}$ and $\pi _{B}$ are measured by 
standard tomographic setup, indicated by $D_\pi$. $BS$ and $D_\pi$ outputs are indicated by $s_j=0,1$,
where the index $j$ refers to the corresponding degrees of freedom.
Hadamard gates $H_{0,1}$ are realized by half wave plates. FF correction apparatus (used only with ordering a)) is given by
a $35m$ length single mode fiber and two Pockels cells ($\sigma_x$, $\sigma_z$) 
driven by the output signals of detector $a_1$, $a_3$, $a_4$.
Dashed lines for $H_{0,1}$, $BS$ and FF correction apparatus indicate that 
these devices can be inserted or not in the setup
depending on the measurement. Inset:
spatial mode matching on the BS.}
\label{fig:schema}
\end{figure}

By the correspondence $\ket H\leftrightarrow\ket0$, $\ket V\leftrightarrow\ket1$, 
$\ket\ell\leftrightarrow\ket0$, $\ket r\leftrightarrow\ket1$, the state $\ket{C_{4}}$ is equivalent to the 
cluster state $\cluster=\frac{1}{2}(\ket+_1\ket0_2\ket0_3\ket+_4+
\ket+_1\ket0_2\ket1_3\ket-_4+\ket-_1\ket1_2\ket0_3\ket+_4-\ket-_1\ket1_2\ket1_3\ket-_4)$ 
\footnote{The state $\cluster$ is obtained by preparing a chain
of qubits all prepared in the state $\ket+$ and then applying the gate 
$CP=|0\rangle \langle 0|\otimes \openone+|1\rangle \langle 1|\otimes \sigma
_{z}$ for each link.} (with $|\pm \rangle =\frac{1}{\sqrt{2}}(|0\rangle \pm
|1\rangle )$) up to single qubit unitaries: 
\begin{equation}
|C_{4}\rangle =U_{1}\otimes U_{2}\otimes U_{3}\otimes U_{4}|\Phi _{4}^{\text{lin}}\rangle \equiv \mathcal U|\Phi _{4}^{\text{lin}}\rangle \,.  
\label{equiv cluster}
\end{equation}
Here $|\Phi _{4}^{\text{lin}}\rangle $ and $|C_{4}\rangle $ are respectively
expressed in the so called ``computational'' and ``laboratory'' basis, while
the $U_{j}$'s ($j=1,\cdots ,4$) are products of Hadamard gates $H=\frac{1}{%
\sqrt{2}}(\sigma _{x}+\sigma _{z})$ and Pauli matrices $\sigma _{i}$. Their
explicit expressions depend on the ordering of the four physical qubits,
namely ${\mathbf{k}}_{A}$, ${\mathbf{k}}_{B}$, $\pi _{A}$, $\pi _{B}$. In
this work we use four different ordering: 
\begin{equation}
\begin{aligned} a)& \text{(1,2,3,4)=}({{\kk}}_B,{{\kk}}_A,\pi_A,\pi_B),
\mathcal U=\sigma_xH\otimes\sigma_z\otimes\openone\otimes H\\ b)&
\text{(1,2,3,4)=}(\pi_B,\pi_A,{{\kk}}_A,{{\kk}}_B), \mathcal
U=H\otimes\sigma_z\otimes\sigma_x\otimes \sigma_zH\\ c)&
\text{(1,2,3,4)=}({{\kk}}_A,{{\kk}}_B,\pi_B,\pi_A), \mathcal
U=\sigma_zH\otimes\sigma_x\otimes\openone\otimes H\\ d)&
\text{(1,2,3,4)=}(\pi_A,\pi_B,{{\kk}}_B,{{\kk}}_A), \mathcal
U=H\otimes\openone\otimes\sigma_x\otimes\sigma_z H. \end{aligned}  \nonumber
\end{equation}
In the following we refer to these expressions depending on the logical
operation we consider.

The general measurement apparatus, differently used to perform each operation,
is sketched in fig. \ref{fig:schema} (see caption for details). 
The $\mathbf{k}$ modes corresponding to photons A or B, are respectively matched on the
up and down side of a common 50:50 beam splitter (BS) (see inset).

\textbf{Single qubit rotations}. In the one-way model a three-qubit linear
cluster state is sufficient to realize arbitrary single qubit rotations \cite{03-rau-mea}.
According to the measurement basis for a generic qubit $j$, 
$|\varphi _{\pm }\rangle _{j}=\frac{1}{\sqrt{2}}(|0\rangle _{j}\pm e^{-i\varphi }|1\rangle _{j})$, we define $s_{j}=0$ 
($s_{j}=1$) when the $|\varphi _{+}\rangle _{j}$ ($|\varphi _{-}\rangle _{j}$)
outcome is obtained. With the 4-qubit cluster in the computational basis ($\cluster$) 
any arbitrary single qubit rotation expressed as 
$|\chi_{out}\rangle =R_{x}(\beta )R_{z}(\alpha )|\chi _{in}\rangle $ 
\footnote{
Three sequential rotations are necessary to implement a generic $SU(2)$
matrix but only two, namely $R_{x}(\beta )R_{z}(\alpha )$, are sufficient to
rotate the input state $|\chi _{in}\rangle =|\pm \rangle $ into a generic state
} 
can be obtained by a three step procedure (cfr. \ref{fig:cluster}(a)).
The sequence of the measurement bases for the three qubits are the following:
I) $\{\ket{0}_1,\ket{1}_1\}$, which allows to obtain a three-qubit linear cluster; 
II) $\ket{\alpha_{\pm}}_2$; III) $\ket{\beta_{\pm}}_3$ or $\ket{-\beta_{\pm}}_3$, 
depending on the second qubit output $s_2=0$ or $s_2=1$ respectively.
In the previous expression $R_{z}(\alpha )=\exp \left( -\frac{i}{2}\alpha \sigma _{z}\right) $,
$R_{x}(\beta )=\exp \left( -\frac{i}{2}\beta \sigma _{x}\right)$ and $\ket{\chi_{in}}=\ket+(\ket-)$ 
if the output of the first measurement is $\ket{0}_1(\ket{1}_1)$.
The output state $\ket{\chi_{out}}$ is created up to Pauli errors ($\sigma _{x}^{s_{3}}\sigma _{z}^{s_{2}}$), that should be
corrected by proper FF operations for deteministic QC \cite{07-pre-hig}.

Let's consider ordering a), where qubit 1=${\kk}_B$, qubit 2=${\kk}_A$, 
qubit 3=$\pi_A$ and qubit 4=$\pi_B$. The output state, encoded in the polarization of photon B, 
can be written in the laboratory basis as 
\eq
|\chi _{out}\rangle_{\pi _{B}}={\sigma _{z}}^{s_{\pi_A}}{\sigma _{x}}^{s_{{\kk}_A}}HR_{x}(\beta)
R_{z}(\alpha )|\chi _{in}\rangle \,,
\fine
where the $H$ gate derives from the change between the computational and laboratory basis. This also implies
that the actual measurement bases are $|\pm \rangle _{\mathbf{k}_{B}}$ for
the momentum of photon B (qubit 1) and $|\alpha _{\mp }\rangle _{\mathbf{k}_{A}}$ for
the momentum of photon A (qubit 2). The measurement basis on the third qubit ($\pi _{A}$) 
depends, according to the one-way model, on the results of the
measurement on the second qubit ($\mathbf{k}_{A}$). These are precisely what 
we call FF measurements (type \textit{i)}), 
corresponding in this scheme to measure 
$\pi _{A}$ in the bases $|\beta _{\pm }\rangle
_{\pi _{A}}$ or $|-\beta _{\pm }\rangle _{\pi _{A}}$, depending on the BS
output mode (i.e. $s_{\kk_A}=0$ or $s_{\kk_A}=1$). Note that these deterministic 
FF measurements directly derive from 
the possibility of encoding two qubits ($\mathbf{k}_{A}$ and $\pi _{A}$) in
the same photon. Differently from the case of 4-photon cluster states, this
avoids the need of Pockels cells to perform active FF
measurements, as already said. 

Pauli errors FF corrections are in any case
necessary for deterministic one way QC. They were realized by using the
measurement apparatus shown in fig. \ref{fig:schema}. Here two fast
driven transverse $LiNbO_{3}$ Pockels cells ($\sigma_x$ and $\sigma_z$)
with risetime $=1n\sec $ and $V_{\frac{\lambda }{2}}\thicksim 1KV$ 
are activated by the output signals of detectors $a_i$ ($i=1,3,4$) corresponding to the different
values of $s_{\pi_A}$ and $s_{\kk_A}$.
They perform the operation ${\sigma _{z}}^{s_{\pi_A}}{\sigma _{x}}^{s_{\kk_A}}$ on photon B, coming
from the output $s_{\kk_B}=0$ of $BS$ and transmitted through a
single mode optical fiber. Note that no correction is needed when
photon A is detected on the output $a_2$ ($s_{\pi_A}=s_{\kk_A}=0$). 
Temporal synchronization between the activation of the high voltage 
signal and the transmission of photon B through the Pockels cells is guaranteed 
by suitable choice of the delays $D$. We used only one $BS$ output of photon $B$, namely $s_{\kk_B}=0$,
in order to perform the algorithm with initial state $\ket{\chi_{in}}=\ket+$. 
The other $BS$ output corresponds to the algorithm starting with the initial state
$\ket{\chi_{in}}=\ket-$.
\begin{figure}[t]
\begin{center}
\includegraphics[width=8cm,height=6cm]{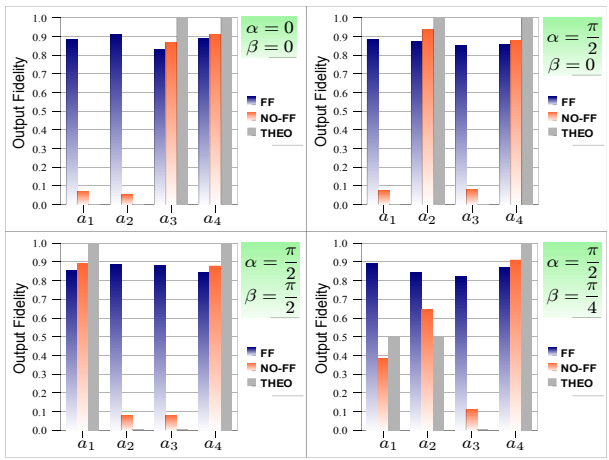}
\end{center}
\caption{Outup fidelities of the single qubit rotation algorithm with (blue columns) or without (red columns) 
feed-forward (FF). In both cases, the four columns of the hystogram refer to the measurement of
the output state (encoded in the polarization of photon B) by detector $b_1$ in
coincidence with $a_1,\dots,a_4$ respectively. 
Grey columns correspond to theoretical fidelities in the no-FF case.
}
\label{fig:risultati}
\end{figure}

In fig. \ref{fig:risultati} the values of the output state fidelity obtained 
with/without active FF corrections
(i.e. turning on/off the Pockels cells) 
are compared for different values of $\alpha$ and $\beta$. 
The expected theoretical fidelities in the no-FF case are also shown . 
In all the cases the computational errors are corrected by the FF action,
with average measured fidelity $F=0.867\pm0.018$. The overall repetition rate is about $500Hz$, 
which is more than 2 orders of magnitude larger than 
one-way single qubit-rotation realized 
with 4-photon cluster states.

We demonstrated the computational equivalence of the two degrees of 
freedom of photon $B$ by performing the same algorithm with ordering b). In this case the 
explicit expression of the output state $|\chi_{out}\rangle_{\mathbf{k}_B}$ 
in the laboratory basis is 
$|\chi_{out}\rangle_{\mathbf{k}_B}={(\sigma_z)}^{s_3}{(\sigma_x)}%
^{s_2}\sigma_zHR_x(\beta)R_z(\alpha)|\chi_{in}\rangle\,$.
By using only detectors $a_2$, $a_3$, $b_1$, $b_2$ in fig. \ref{fig:schema} we measured $|\chi_{out}\rangle_{%
\mathbf{k}_B}$ choosing different values of $\alpha$ (which correspond in
the laboratory to the polarization measurement bases $|\alpha_{\mp}\rangle_{\pi_A}$) and 
$\beta=0$ (which correspond in the laboratory to the momentum bases $%
|-\beta_{\pm}\rangle_{\mathbf{k}_A}$). The first qubit ($\pi_B$) was
always measured in the basis $|\pm\rangle_{\pi_B}$. By 
performing the $\mathbf{k}_B$ tomographic analysis for all the possible 
values of $s_2\equiv s_{\pi_A}$ and $s_3\equiv s_{\kk_A}$ and of the input qubit (i.e. for different
values of $s_1\equiv s_{\pi_B}$) we obtained
an average value of fidelity $F>0.9$ (see table \ref{table:rotation}).
In this case the realization of FF corrections could be realized by the adoption of active phase modulators.
The ($\pi $)-(${\mathbf{k}}$) computational equivalence and the use of active 
feed-forward show that the multidegree of freedom approach is feasible for deterministic one-way QC.
\begin{table}
	\begin{ruledtabular}
				\centering
			\begin{tabular}{c|c|c}
			$\alpha (\beta=0)$ & F $(s_2=s_3=0)$ & F $(s_2=0,s_3=1)$ 
			\\
			\hline
			\hline
			$0$ & $0.961\pm0.003$ & $0.971\pm0.003$ 
			\\
			\hline
			$\pi/2$ & $0.879\pm0.006$ & $0.895\pm0.005$ 
			\\
			\hline
			$\pi/4$ & $0.998\pm0.005$ & $0.961\pm0.006$ 
			\\	
			\hline
			$-\pi/4$ & $0.833\pm0.007$ & $0.956\pm0.006$ 
\end{tabular}
\end{ruledtabular}
		\caption{Momentum (${\bf k}_B$) experimental fidelities (F) 
		of single qubit rotation output states for different values of $\alpha$ and $\beta=0$. 
		Each datum is obtained by the measurements of the different Stokes parameters, each one lasting 10 sec.}
\label{table:rotation}
\end{table}

{\bf C-NOT gate for equatorial qubits}. 
Nontrivial two-qubit operations, such as the C-NOT gate, can be realized by the four-qubit horseshoe (180$^\circ$ rotated) 
cluster state (see fig. \ref{fig:cluster}b)),
whose explicit expression is equal to $\cluster$.
By simultaneously measuring qubits 1 and 4, it's possible to implement the logical circuit shown
in fig. \ref{fig:cluster}(b). In the computational basis the input state is $\ket{+}_c\otimes\ket{+}_t$ (c=control, t=target),
while the output state, encoded in qubits 2 (control) and 3 (target),
is $\ket{\Psi_{out}}=H_tC\text{-}NOT(\mathcal O\ket{+}_c\otimes R_z(\alpha)\ket{+}_t)$
(for $s_1=s_4=0$).
In the above expression we have $\mathcal O=\openone$ ($\mathcal O=H$) when qubit 1 is measured in the basis $\{\ket0_1,\ket1_1\}$
($\ket{\pm}_1$). Qubit 4 is measured in the basis $\ket{\alpha_{\pm}}_4$.
It is worth noting that this circuit realizes the C-NOT gate (up to the Hadamard $H_t$)
for arbitrary equatorial target qubit and control qubit $\ket0,\ket1$ or $\ket{\pm}$ depending on the
measurement basis of qubit 1.
\begin{table}[t]
		\begin{ruledtabular}
						\centering
		\begin{tabular}{c|c|c|c|c}
			$\quad\mathcal O\quad$ & $\alpha$ & $\text{Control output}$ &  $F(s_4=0)$ & $F(s_4=1)$
			\\
			\hline\hline			
			             &   $\pi/2$ & $s_1=0\rightarrow\ket1_c$  & $0.965\pm0.004$ & $0.975\pm0.004$
			\\
			\cline{3-5}
			$H$          &    &$s_1=1\rightarrow\ket0_c$  & $0.972\pm0.004$ & $0.973\pm0.004$
			\\
			\cline{2-5}
			& $\pi/4$  &$s_1=0\rightarrow\ket1_c$  & $0.995\pm0.008$ & $0.902\pm0.012$
			\\
			\cline{3-5} 
      &			&$s_1=1\rightarrow\ket0_c$   & $0.946\pm0.010$ & $0.945\pm0.009$
			\end{tabular}
			\begin{tabular}{c|c|c|c|c}
  		$\ \mathcal O\ $ & $\alpha$ & $\text{Control output}$ &  $F(s_1=s_4=0)$ & $F(s_1=0,s_4=1)$
	 		\\
	 		\hline\hline
  		&  $\pi/2$  &$\ket0_c\equiv\ket \el_{{\bf k}_B}$  & $0.932\pm0.004$ & $0.959\pm0.003$
			\\
			\cline{3-5}
			$\openone$&&$\ket1_c=\ket r_{{\bf k}_B}$    & $0.941\pm0.005$ & $0.940\pm0.005$
			\\
			\cline{2-5} 
			&  $\pi/4$  &$\ket0_c=\ket \el_{{\bf k}_B}$  			& $0.919\pm0.007$ & $0.932\pm0.007$
			\\
			\cline{3-5}
			& &$\ket1_c=\ket r_{{\bf k}_B}$    	& $0.878\pm0.009$ & $0.959\pm0.006$
		\end{tabular}
		\end{ruledtabular}
		\caption{Experimental fidelity (F) of C-NOT gate output target qubit for different value of $\alpha$ and $\mathcal O$.} 
		\label{table:c-not}
\end{table}

The experimental realization of this gate was performed by adopting ordering c).
In this case the control output qubit is encoded in the momentum ${\bf k}_B$, while the target
output is encoded in the polarization $\pi_B$. 
In the actual experiment we inserted $H_0$ and $H_1$ on photon B to compensate $H_t$.
The output state in the laboratory basis is then
\eq
\ket{\Psi_{out}}=(\Sigma)^{s_4}\sigma^{(c)}_x
C\text{-}NOT(\mathcal O\sigma^{s_1}_z\ket{+}_c\otimes R_z(\alpha)\ket{+}_t)\,,
\fine
where all the possible measurement outcomes of qubits 1 and 4 are considered. The Pauli errors are
$\Sigma=\sigma^{(c)}_z\sigma^{(t)}_z$, while the matrix $\sigma^{(c)}_x$ is due to the change
between the computational and laboratory bases.
Table \ref{table:c-not} shows the experimental fidelities of the target qubit
corresponding to the measurement of the output control qubit in the basis $\{\ket0,\ket1\}$.

\textbf{Universal C-Phase gate}. We realized a C-Phase gate for arbitrary target qubit
and fixed control $|+\rangle_c$ (see Fig. \ref{fig:cluster}(c)) by measuring 
qubits 1 and 2 of $\cluster$ in the bases 
$|\alpha_\pm\rangle$ and $\ket{(-)^{s_1}\beta_\pm}$ respectively. 
By considering ordering d) we encoded the output state in
the physical qubits $\mathbf{k}_A$ and $\mathbf{k}_B$. For $s_1=s_2=0$, by
using the appropriate base changing, the output state is written as 
\begin{equation}
|\Psi_{out}\rangle=|-\rangle_{\mathbf{k}_A}\otimes\sigma_x|\Phi\rangle_{%
\mathbf{k}_B}+|+\rangle_{\mathbf{k}_A}\otimes\sigma_x\sigma_z|\Phi\rangle_{%
\mathbf{k}_B}\,.
\end{equation}
Here $\ket{\Phi}_{{\bf k}_B}=R_x(\beta)R_z(\alpha)\ket+$ and the matrix $\sigma_x$ is due to the basis changing.
We measured the fidelity of target ${\bf k}_B$ corresponding to a control $\ket+_{{\bf k}_A}$ ($\ket-_{{\bf k}_A}$)
for different values of $\alpha$ and $\beta$, obtaining an average value $F=0.907\pm0.010$ ($F=0.908\pm0.011$).

In this Letter 2-photon 4-qubit cluster states, realizing the full 
entanglement of two photons by two degrees of freedom (in our 
case polarization and linear momentum), have been used to perform 
highly efficient arbitrary single qubit rotations, either probabilistic or 
deterministic, and fundamental two qubit gates, such as a C-NOT gate for 
target qubits located in the equatorial plane of the Bloch sphere 
and a C-Phase gate for generic target qubits. These operations
have been performed with high values of fidelity and at average 
repetition rates which are 2 or even 3 orders of magnitude larger 
than those obtained with 4-photon cluster states.  

The power of computation is strongly related to the possibility of 
increasing the information content associated to a quantum state.
For instance, 6 qubits are necessary to implement a C-NOT gate 
operating over the entire Bloch sphere of the target and control qubits. 
This could be obtained by using other degrees of freedom of the photon, 
such as time-energy. QC based on more complex gates and algorithms requires to work 
with even more qubits \cite{03-rau-mea}. In our scheme this could be 
realized by exploiting the SPDC conical emission of a type I crystal 
and using a larger number of {\bf k} modes. Even if this number scales 
exponentially with the number of qubits, up to eight qubits 
could be created with two photons by using only four modes per photon, besides polarization and time-bin. 
At the same time, complex QC operations can not be realized without 
the availability of a larger number of photon pairs. This number should 
grow contextually with the number of available 
degrees of freedom of the photon. For instance eight-qubit four-photon 
cluster states could be generated by linking together two $|C_4\rangle$ 
states by a proper CP gate. 
This leads to conceive a hybrid approach to one-way QC based on a multiphoton 
multiqubit architecture, which is at the moment under investigation. 

\begin{acknowledgments}
Thanks are due to Fabio Sciarrino for useful discussions and Marco Barbieri for
his contribution in planning the experiment. This work was supported by the 
contract PRIN 2005 of MIUR (Italy).
\end{acknowledgments}


\end{document}